\documentclass[twocolumn,showpacs,preprintnumbers,amsmath,amssymb]{revtex4}
\usepackage{graphicx}% Include figure files
\usepackage{dcolumn}% Align table columns on decimal point
\usepackage{bm}% bold math

\newcommand{\GeV}{\ensuremath{\text{GeV}}}
\newcommand{\MeV}{\ensuremath{\text{MeV}}}
\newcommand{\bgs}{\ensuremath{B\to X_s\gamma}}
\newcommand{\ifb}{\ensuremath{\,\rm fb^{-1}}}

\begin{document}

\preprint{BELLE-CONF-0802}

%\title{Moments of the photon energy spectrum and partial branching \\ fraction of $B\to X_s\gamma$ decays at Belle.}% Force line breaks with \\
\title{Improved Measurement of Inclusive Radiative $B$-meson decays}% Force line breaks with \\

%%% Paper:    
%%% Journal:  Summer 2007 Conference Papers
%%% August 23, 2007
%%% Contacts: 
%%% Non-responding authors or those who said NO are commented out.
%%% ====================================================================
%%% Click the RELOAD button on your web browser to see the updated file.
%%% ====================================================================
%%% Use \input{author} to insert this material into your latex file.
%%%%% Force institutions to appear in alphabetical order when typeset.
\affiliation{Budker Institute of Nuclear Physics, Novosibirsk}
\affiliation{Chiba University, Chiba}
\affiliation{University of Cincinnati, Cincinnati, Ohio 45221}
\affiliation{Department of Physics, Fu Jen Catholic University, Taipei}
\affiliation{Justus-Liebig-Universit\"at Gie\ss{}en, Gie\ss{}en}
\affiliation{The Graduate University for Advanced Studies, Hayama}
\affiliation{Gyeongsang National University, Chinju}
\affiliation{Hanyang University, Seoul}
\affiliation{University of Hawaii, Honolulu, Hawaii 96822}
\affiliation{High Energy Accelerator Research Organization (KEK), Tsukuba}
\affiliation{Hiroshima Institute of Technology, Hiroshima}
\affiliation{University of Illinois at Urbana-Champaign, Urbana, Illinois 61801}
\affiliation{Institute of High Energy Physics, Chinese Academy of Sciences, Beijing}
\affiliation{Institute of High Energy Physics, Vienna}
\affiliation{Institute of High Energy Physics, Protvino}
\affiliation{Institute for Theoretical and Experimental Physics, Moscow}
\affiliation{J. Stefan Institute, Ljubljana}
\affiliation{Kanagawa University, Yokohama}
\affiliation{Korea University, Seoul}
\affiliation{Kyoto University, Kyoto}
\affiliation{Kyungpook National University, Taegu}
\affiliation{\'Ecole Polytechnique F\'ed\'erale de Lausanne (EPFL), Lausanne}
\affiliation{University of Ljubljana, Ljubljana}
\affiliation{University of Maribor, Maribor}
\affiliation{University of Melbourne, School of Physics, Victoria 3010}
\affiliation{Nagoya University, Nagoya}
\affiliation{Nara Women's University, Nara}
\affiliation{National Central University, Chung-li}
\affiliation{National United University, Miao Li}
\affiliation{Department of Physics, National Taiwan University, Taipei}
\affiliation{H. Niewodniczanski Institute of Nuclear Physics, Krakow}
\affiliation{Nippon Dental University, Niigata}
\affiliation{Niigata University, Niigata}
\affiliation{University of Nova Gorica, Nova Gorica}
\affiliation{Osaka City University, Osaka}
\affiliation{Osaka University, Osaka}
\affiliation{Panjab University, Chandigarh}
\affiliation{Peking University, Beijing}
\affiliation{University of Pittsburgh, Pittsburgh, Pennsylvania 15260}
\affiliation{Princeton University, Princeton, New Jersey 08544}
\affiliation{RIKEN BNL Research Center, Upton, New York 11973}
\affiliation{Saga University, Saga}
\affiliation{University of Science and Technology of China, Hefei}
\affiliation{Seoul National University, Seoul}
\affiliation{Shinshu University, Nagano}
\affiliation{Sungkyunkwan University, Suwon}
\affiliation{University of Sydney, Sydney, New South Wales}
\affiliation{Tata Institute of Fundamental Research, Mumbai}
\affiliation{Toho University, Funabashi}
\affiliation{Tohoku Gakuin University, Tagajo}
\affiliation{Tohoku University, Sendai}
\affiliation{Department of Physics, University of Tokyo, Tokyo}
\affiliation{Tokyo Institute of Technology, Tokyo}
\affiliation{Tokyo Metropolitan University, Tokyo}
\affiliation{Tokyo University of Agriculture and Technology, Tokyo}
\affiliation{Toyama National College of Maritime Technology, Toyama}
\affiliation{Virginia Polytechnic Institute and State University, Blacksburg, Virginia 24061}
\affiliation{Yonsei University, Seoul}
  \author{K.~Abe}\affiliation{High Energy Accelerator Research Organization (KEK), Tsukuba} % KEK
  \author{I.~Adachi}\affiliation{High Energy Accelerator Research Organization (KEK), Tsukuba} % KEK
  \author{H.~Aihara}\affiliation{Department of Physics, University of Tokyo, Tokyo} % Tokyo
  \author{K.~Arinstein}\affiliation{Budker Institute of Nuclear Physics, Novosibirsk} % BINP
  \author{T.~Aso}\affiliation{Toyama National College of Maritime Technology, Toyama} % Toyama
  \author{V.~Aulchenko}\affiliation{Budker Institute of Nuclear Physics, Novosibirsk} % BINP
  \author{T.~Aushev}\affiliation{\'Ecole Polytechnique F\'ed\'erale de Lausanne (EPFL), Lausanne}\affiliation{Institute for Theoretical and Experimental Physics, Moscow} % ITEP
  \author{T.~Aziz}\affiliation{Tata Institute of Fundamental Research, Mumbai} % Tata
  \author{S.~Bahinipati}\affiliation{University of Cincinnati, Cincinnati, Ohio 45221} % Cincinnati
  \author{A.~M.~Bakich}\affiliation{University of Sydney, Sydney, New South Wales} % Sydney
% ...
% Lausanne
\collaboration{The Belle Collaboration}
\noaffiliation

%\author{Antonio Limosani (Belle Collaboration)}

\begin{abstract}
% abstract goes here
We report a  fully inclusive measurement of the
flavor changing neutral current decay \bgs\ in the
energy range $1.7\,\GeV\le E^\mathrm{c.m.s}_\gamma\le2.8\,\GeV$, covering
97\% of the total spectrum, where c.m.s is the center of mass system.
Using $605\ifb$ of data, we obtain in the rest frame of the $B$-meson ${\mathcal B}(B\to
X_s\gamma : E^B_\gamma>1.7\,\mathrm{GeV})= \left(3.31 \pm 0.19 \pm 0.37 \pm 0.01\right)\times 10^{-4}$,
where the errors are statistical, systematic and from the boost
correction needed to transform from the rest frame of the $\Upsilon(4S)$ (c.m.s) to that of the $B$-meson, respectively.
We also measure the first and second
moments of the photon energy spectrum as functions of various energy
thresholds, which extend down to $1.7\,\GeV$. The results are preliminary.
\end{abstract}

\pacs{12.39.Hg, 13.20.He, 13.40.Hq, 14.40.Nd, 14.65.Fy} 
% 12.39.Hg Phenomenological quark models - Heavy quark effective theory
% 13.20.He Leptonic, semileptonic, and radiative decays of mesons - Decays of bottom mesons
% 13.40.Hq Electromagnetic processes and properties - Electromagnetic decays
% 14.65.Fy Properties of specific particles - Quarks - Bottom quarks
% 14.40.Nd Properties of specific particles - Mesons - Bottom mesons
\maketitle

%\setcounter{footnote}{0}

% body of text goes here

Radiative $B$-meson decays may offer a view of phenomena beyond the
Standard Model of particle physics (SM). In the SM, these decays proceed
via a flavor changing neutral current (FCNC) decay, which consists of
a loop process.
%as depicted Fig~\ref{fig:Feynman}
Yet to be discovered
particles, such as charged Higgs or supersymmetric particles, may be
produced virtually in the loop and produce a measureable deviation
from the branching fraction predicted by the SM.

The predictions of the branching fraction at order $\alpha_s^2$ (NNLO - next to next to leading
order)
$\left(3.15 \pm 0.23\right)\times10^{-4}$~\cite{Misiak:2006zs}, 
$\left(2.98 \pm 0.26\right)\times10^{-4}$~\cite{Becher:2006pu}  
%$\left(3.57 \pm 0.39\right)\times10^{-4}$~\cite{Andersen:2006hr}
and the average of experiment measured values $\left(3.55\pm0.26\right)\times10^{-4}$~\cite{Yao:2006px}
are in tacit agreement. An updated experimental measurement would
further test this agreement, and, moreover, give stronger constraints
on extensions to the SM {\it e.g.}  Minimal Supersymmetric Standard
Model~\cite{Bertolini:1990if} and left-right symmetric
models~\cite{Cho:1993zb,Fujikawa:1993zu}.

The photon energy spectrum is
also of great importance. At the parton level, the photon is monochromatic with energy 
$E\approx m_b/2$ in the $b$-quark rest frame. 
The energy is smeared by the motion of the $b$-quark inside the $B$
meson and gluon emission.
A measurement of the moments of this spectrum allows for a 
determination of the $b$-quark mass and of its {\it Fermi motion}. 
This information can then be used in the extraction of the CKM matrix elements 
$|V_{cb}|$ and $|V_{ub}|$ from inclusive 
semileptonic $B$ decays~\cite{Barberio:2007cr}. A measurement of the low-energy tail of the photon
spectrum is important in this context~\cite{Bigi:2002qq}.

Belle has previously measured the \bgs\ branching fraction
with $5.8\ifb$ and $140\ifb$ of data using semi-inclusive~\cite{Abe:2001hk} and fully
inclusive approaches~\cite{Koppenburg:2004fz}, respectively.
%In the former measurement
%we required $E_\gamma>1.8\,\GeV$ in the $\Upsilon(4S)$ rest
%frame~\cite{CM}.
Other measurements include those
from CLEO~\cite{Chen:2001fja} and BaBar~\cite{Aubert:2005cua,Aubert:2006gg,Aubert:2007my}.

Here we present an update of our fully inclusive measurement~\cite{Koppenburg:2004fz},
based on a much larger dataset and with
significant refinements, which includes an unfolding of detector effects on the
measured spectrum that improve the measurements of the
branching fraction and spectral moments, respectively. We also
extend the photon energy range to
$E^{\mathrm{c.m.s}}_\gamma>1.7\,\GeV$,
covering more of the spectrum than ever before,  where c.m.s refers
to the centre of mass system, which is equivalent to the rest frame of
the $\Upsilon(4S)$.

The \bgs\ decay is studied using the Belle detector at the KEKB
asymmetric $e^+e^-$ storage ring~\cite{KEKB}. The data consists a
of sample of $604.6\ifb$ taken at the $\Upsilon(4S)$ resonance
corresponding to $(656.7\pm8.9)\times10^6$ $B\bar{B}$ pairs. Another
$68.3\ifb$ sample has been taken at an energy $60\,\MeV$ below the resonance 
and is used to measure the non-$B\bar{B}$ background.
Throughout this manuscript, we refer to these data samples as the ON and OFF
samples, respectively.

% From Zhang, Knpi
%
% The Belle detector is a large-solid-angle magnetic spectrometer.
% The main component relevant for this analysis is the 
% electromagnetic calorimeter (ECL) located inside the magnet. It is made
% of 8736 CsI(Tl) crystals of $5.5\times 5.5\rm\,cm^2$ cross-section at the
% front surface and $16.2$ radiation lengths long. The photon energy resolution
% is approximately 2\% for the energy range relevant in this analysis.
% Charged particle tracking is provided by a three-layer silicon vertex
% detector and a 50-layer central drift chamber (CDC), both located in 
% a $1.5\,\rm T$ magnetic field. Electron identification is performed 
% using the shower shape information in the ECL, an aerogel \v{C}erenkov
% counter (ACC) and the $dE/dx$ measurement in the CDC. Muons are identified
% using the instrumented iron flux-return located outside of the coil (KLM).
% The detector is described in detail elsewhere~\cite{Belle}.
The Belle detector is a large-solid-angle magnetic spectrometer
described in detail elsewhere~\cite{NIM:Belle}.
The main component relevant for this analysis is the 
electromagnetic calorimeter (ECL) made of 
$16.2$ radiation lengths long CsI(Tl) crystals. The photon energy resolution
is about 2\% for the energy range relevant in this analysis.

% STRATEGY

The strategy to extract the signal \bgs\ spectrum is to collect all 
high-energy photons, vetoing those originating from $\pi^0$ and $\eta$ 
decays to two photons. The contribution from continuum $e^+e^-\to q\bar{q}$
($q=u,d,s,c$) and QED type events is subtracted using the OFF sample.
The remaining backgrounds from $B\bar{B}$ events are subtracted using 
Monte Carlo (MC) distributions scaled by data control samples.

% Photon selection

Photon candidates are selected from ECL clusters of $5\times5$ crystals
in the barrel region ($-0.35\le\cos\theta\le0.70$, where $\theta$ 
is the polar angle with respect to the beam axis, subtended from the
direction opposite the positron beam. 
They are required to have an energy 
$E^\mathrm{c.m.s}_\gamma$ larger than $1.4\,\GeV$. We require $95$\% of the energy to
be deposited in the central $3\times3$ crystal array and use
isolation cuts to veto photons from bremsstrahlung and interaction with matter.
The center of the cluster has to be displaced
from any other ECL cluster with $E>20\,\MeV$ by at least $30\,\rm cm$ 
at the surface of the calorimeter, 
and from any reconstructed track by $30\,\rm cm$,
or by $50\,\rm cm$ for tracks with a measured momentum above $1\,\GeV/c$. 
Moreover, the angle between the photon and the highest energy lepton
in the event has to be larger than $0.3$ radians at the interaction point.

In the Belle detector, a non-negligible background (1\%)
is due to the overlap of a hadronic event
with energy deposits left in the calorimeter by previous 
QED interactions (mainly Bhabha scattering). 
Such composite events are completely removed
using timing information for calorimeter clusters 
associated with the candidate photons.
The cluster timing information is stored in the raw data,
and is available in the reduced format used for
analysis only for data processed after the summer of 2004. This divides
our data set into $253.7\ifb$ and $350.9\ifb$  samples of
reprocessed data without and
with timing information, respectively. To minimise composite background
due to Bhabha scattering and two-photon processes that
contaminate both $\Upsilon(4S)$ and continuum data samples, we 
veto any candidate that contains an ECL cluster with energy exceeding $1$
GeV within a cone of 0.2 radians in the
direction opposite our photon candidate as measured in the c.m.s frame. In the second data set
only photons that are in time with the rest of the event are 
retained. The efficiency of this selection on signal events
is larger than 99.5\%.
%In our previous analysis, which
%fell in the first set of data this
%and other sources of beam background were modelled using a sample of randomly triggered 
%events. 
We veto candidate photons from $\pi^0$ and $\eta$ decays to two photons by
combining each $B\to X_s\gamma$ candidate photon with all other
photons in the event. We reject the photon candidate if the likelihood
of being a $\pi^0$ or $\eta$ is larger than $0.1$ and $0.2$,
respectively, these yield, on average, background suppression factors of $4$ and
$2$, respectively. These likelihoods are determined from MC and are functions of the 
laboratory energy of the other photon, its polar angle $\theta$ and the mass of the two-photon
system.

In order to reduce the contribution from continuum
events, we use two Fisher discriminants calculated in the c.m.s frame. %~\cite{Fisher}.
The first discriminant exploits the topology of 
\bgs\ events and combines three energy flows around the photon axis.
% From Alan:
% \bgs\ events and combines three variables characterizing energy flow around the photon axis.
% They are obtained using the energy sums of all particles except the candidate 
% photon whose direction lies in the cones 
% - From Tom:
%These energy flow variables are obtained using all particles, except 
%for the photon candidate, whose direction lies in the regions
%$\alpha^\ast<30^\circ$, $\alpha^\ast>140^\circ$, where $\alpha^\ast$ 
%is the angle to the candidate photon, and in the region in between.
% - From Nick Parslow:
These energy flow variables are obtained using all particles, except 
for the photon candidate, we measure the energy in the three regions
defined by $\Theta<30^\circ$, $30^\circ\le\Theta\le140^\circ$,
$\Theta>140^\circ$, where $\Theta$ 
is the angle of the particle to the candidate photon.
The second exploits the spherical shape of $B\bar{B}$ events and 
is built using ten event-shape variables including 
Fox-Wolfram moments~\cite{Fox:1978vu} for the full event and for the partial event
with the photon removed, the full- and partial-event thrusts and the
angles of the thrust axis with respect to the beam and the photon direction.
% - From Tom:
%is built using ten event-shape variables.
%These variables are calculated using either all tracks and showers
%in the event or excluding the photon candidate. The event
%shape variables include Fox-Wolfram moments, thrust, and the angles
%of the thrust axis with respect to the beam and photon direction.
% To optimize these selection criteria, we use a MC simulation~\cite{MC} 
% containing large samples of $B\bar{B}$, $q\bar{q}$ and signal weighted
% according to the expected number of events in the considered ON and OFF 
% samples. 
% - Tom:
To optimise these selection criteria, we use a MC
simulation~\cite{Brun:1978fy} containing large samples of $B\bar{B}$, $q\bar{q}$ and
signal weighted according to the luminosities of the ON and OFF
samples. In the optimisation step the signal MC used is generated as
inclusive $B\to X_s\gamma$ and exclusive $B\to K^*\gamma$. The
inclusive component $X_s$ is defined as a resonance of
spin-1 with a Breit-Wigner form and a mass of
$2.4$ GeV/$c^2$ and width $1.5$ GeV/$c^2$. The $X_s$ system is
hadronised by JETSET
and subsequently reweighted to match the prediction of the DGE model~\cite{Andersen:2005bj}~\footnote{In the
  optimisation step the choice of signal model has a negligible
  effect on the measure of optimisation, suffice to say the choice of
  signal model should not be construed as preferential.}
with $m_b(\overline{\mathrm{MS}})=4.20\,\mathrm{GeV}/c^2$, with the mass
extending no lower than $1.18\,\mathrm{GeV}/c^2$ to agree with the
corresponding world
average branching fractions~\cite{Yao:2006px}.
To improve the understanding of the photon energy spectrum 
at low energies, the selection criteria are
optimised to maximize the sensitivity to the signal in
the energy bin $1.8\,\GeV<E^\mathrm{c.m.s}_\gamma<1.9\,\GeV$.

%The signal MC is generated as a weighted sum
%of $B\to K^\ast\gamma$ decays, where $K^\ast$ is any known spin-$1$ resonance
%with strangeness $S=1$. The relative weights are obtained by fitting 
%the total photon spectrum to a theoretical model~\cite{Kagan}.
%The signal MC is normalized to the average measured
%branching fraction~\cite{Yao:2006px}. 
%The measured energy spectrum is histogrammed in bins of $100\,\MeV$ 
%to comply with previous references~\cite{CLEOb2g}.
%, which is the most critical.
% From this MC study, we expect to observe an excess
% of data of one standard deviation or more in the $100\,\MeV$ bins
% ranging from $1.8$ to $2.8\,\GeV$.

After these selection criteria we observe $4.15\times10^6$ and $0.25\times10^6$ 
photon candidates in the ON and OFF data samples, respectively.
%and $1.1\times10^5$ in OFF data. 
%There is no limit to the number of accepted candidates per event.
The spectrum measured in OFF data is scaled by luminosity to 
the expected number of non-$B\bar B$ events in ON data and subtracted.
The formula used to subtract continuum background is as follows:
\begin{eqnarray*}
N^{B\bar{B}}(E^{\mathrm{c.m.s(ON)}}_\gamma)  & = &
N^{\mathrm{ON}}(E^{\mathrm{c.m.s(ON)}}_\gamma) -  \\
& & \alpha \cdot
\frac{\epsilon^{\mathrm{ON}}_{\mathrm{Hadronic}}}{\epsilon^{\mathrm{OFF}}_{\mathrm{Hadronic}}}
\cdot\frac{\epsilon^{\mathrm{ON}}_{B\to X_s\gamma}}{\epsilon^{\mathrm{OFF}}_{B\to X_s\gamma}}
\cdot F_N \\
& & \cdot N^{\mathrm{OFF}}(F_E E^{\mathrm{c.m.s(OFF)}}_\gamma) \\
\end{eqnarray*}
where $\epsilon$ is the efficiency of Belle's hadronic selection~\cite{Casey} or of this
analysis' ($B\to X_s\gamma$) selection criteria in continuum events at either ON resonance
($\sqrt{s}=10.58\,$ GeV) or OFF resonance ($\sqrt{s}=10.52\,$ GeV)
energies, and $\alpha$ is the ratio of ON to OFF resonance
integrated luminosity corrected for the energy difference ($\alpha=8.7557(\pm
0.3\%)$). The factors $F_E$ and $F_N$ compensate for the slightly lower mean
energy and multiplicity of particles in OFF compared to ON events. We
find $F_N=1.0009 \pm 0.0001$, $F_E=1.0036 \pm 0.0001$, 
$\frac{\epsilon^{\mathrm{ON}}_{\mathrm{Hadronic}}}{\epsilon^{\mathrm{OFF}}_{\mathrm{Hadronic}}}=0.9986 \pm 0.0001$,
and $\frac{\epsilon^{\mathrm{ON}}_{B\to
    X_s\gamma}}{\epsilon^{\mathrm{OFF}}_{B\to X_s\gamma}}=0.9871 \pm
0.0014$. The ON and scaled OFF spectra and their difference are shown in 
Fig.~\ref{fig:Bdata}.

%%%%%%%%%%%%%%%%%%%%%%%%%%%%%%%%%%%%%%%%%%%%%%%%%%%%%%%%%%%%%%%%%%%%%%%%%%%%%%%%%
\begin{figure}
  \begin{tabular}{c}
\hspace{-5mm} \includegraphics[scale=0.44]{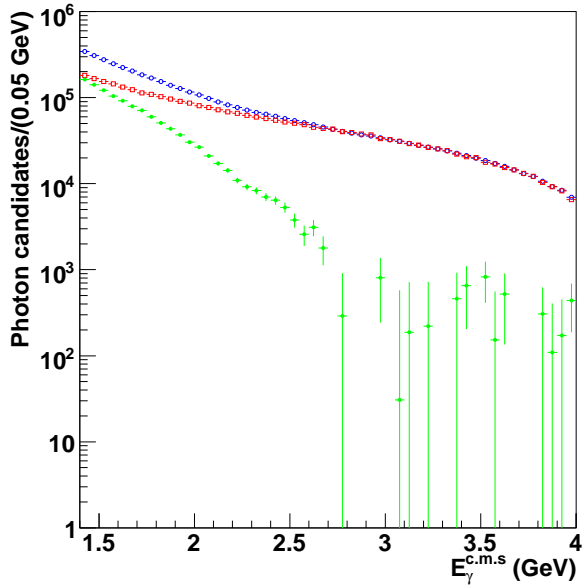} \\
\end{tabular}
\caption{\label{fig:Bdata}
ON data (open circle), scaled OFF data (open square) and
continuum background subtracted (filled circle) photon energy
spectra of candidates in the c.m.s frame.
}
\end{figure}
%%%%%%%%%%%%%%%%%%%%%%%%%%%%%%%%%%%%%%%%%%%%%%%%%%%%%%%%%%%%%%%%%%%%%%%%%%%%%%%%%

%these factors respectively compensate: for the less energy
%available to photon candidates; the lower multiplicity of
%photon candidates; the response to Belle's hadronic event selection
%criteria; and the response to selection criteria of this analysis.
%All factors have been all derived from a MC study.

We then subtract the backgrounds from $B$ decays from the obtained spectrum.
% According to MC, in the $1.8$--$2.8\,\GeV$ range, more than half of this 
% spectrum is due to photons from $\pi^0$ decays.
Six background categories are considered: 
{\it (i)} photons from $\pi^0\to\gamma\gamma$;
{\it (ii)} photons from $\eta\to\gamma\gamma$;
{\it (iii)} other real photons (mainly decays of $\omega$, $\eta'$, and $J/\psi$, 
and bremsstrahlung,
including the short distance radiative correction (modelled with PHOTOS~\cite{Barberio:1993qi});
%including those radiated at the weak
%vertex of semileptonic $B$ decays as modelled with PHOTOS~\cite{Barberio:1993qi});
{\it (iv)} ECL clusters not due to single photons (mainly $K^0_L$'s and $\bar{n}$'s);
{\it (v)} Electrons %interacting with matter
misidentified as photons and;
{\it (vi)} beam background.
The spectra of the background of photons from $B$-meson decays with
respect to the expected signal is shown in Fig.~\ref{fig:Bbkgd} and
listed in Table~\ref{tab:Bbkgd}. The net background of this type is a
factor five greater than the signal.

\begin{table}
  \begin{tabular}{|l|c|}\hline
    Contribution & Fraction \\\hline
    Signal                & 0.190  \\\hline
    Decays of $\pi^0$     & 0.474  \\
    Decays of $\eta$      & 0.163  \\
    Other secondary $\gamma$      & 0.081  \\
    Mis-IDed electrons    & 0.061  \\
    Mis-IDed hadrons      & 0.017  \\
    Beam background       & 0.013  \\\hline
  \end{tabular}
  \caption{Relative contributions of the $B\overline{B}$ backgrounds
  after selection in the $1.7<E^\mathrm{c.m.s}_\gamma/(\mathrm{\,GeV})<2.8$ range \label{tab:Bbkgd}}
\end{table}

\begin{figure}[htb]
  \begin{tabular}{c}
  \includegraphics[scale=0.33]{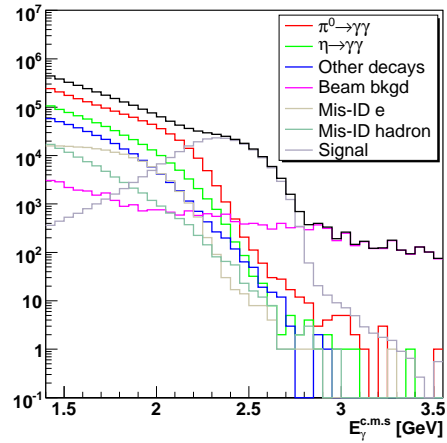} \\
  \end{tabular}
  \caption{The spectra of photons from $B$-meson decays passing
    selection criteria as predicted
    using a MC sample. \label{fig:Bbkgd}}
\end{figure}

For each of these categories we take the predicted background
from MC and scale it according to measured yields wherever possible.
The inclusive $B\to\pi^0X$ and $B\to\eta X$ spectra are measured in data
using pairs of photons with well-balanced energies
and applying the same ON$-$OFF subtraction procedure. 
The yields obtained in data are on average $10$\% larger and $5$\% lower for
$\pi^0$ and $\eta$ %without correction       
than MC expectations.
%depending on the photon energy range.
The observed discrepancy between the measured and simulated $\pi^0$
$\eta$ spectra is attributed to the branching fraction 
assumptions used for the generator~\cite{Lange:2001uf}.
Beam background is measured using a sample of randomly triggered 
events and added to the $B\bar{B}$ MC. 
%Other yields are not scaled at this stage.

For each selection criterion and each background
category we determine the $E^\mathrm{c.m.s}_\gamma$-dependent selection efficiency 
in OFF-subtracted ON data and MC using appropriate control samples. We then
scale the MC background sample according to the ratio
of these efficiencies. 
The efficiencies of the $\pi^0$ and $\eta$ vetoes for photons not from
$\pi^0$ and $\eta$ are measured in data using one photon from a
reconstructed $\pi^0$, where the other photon of the $\pi^0$ is
excluded from the search over the remaining photons for the next best
$\pi^0$ or $\eta$ candidate (highest $\pi^0$ or $\eta$ likelihood).
Consequently the best formed $\pi^0$ or $\eta$ candidate used in the
calculation of the likelihoods is most likely a random combination,
and therefore suited to measuring the effect of the vetoes. 
The $\pi^0$ veto efficiency is measured
using a sample of photons coming from measured $\pi^0$ decays.
We use partially reconstructed $D^{\ast+}\to D^0\pi^+$, $D^0\to K^-\pi^+\pi^0$
decays where the $\pi^0$ is replaced by the candidate photon
in the reconstruction. The $\eta$ veto efficiency for photons from $\pi^0$'s and event-shape criteria efficiencies 
are measured using a $\pi^0$ anti-veto sample, which is made of
photons with a $\pi^0$ likelihood larger than $0.75$ ({\it i.e}, no
$\pi^0$ veto) and passing all other selection criteria.
Other efficiencies are measured using the signal sample. Beam
background is negligible after the application of the OFF time veto.
In the sample of data where the veto is unavailable we scale the
background according to a comparison of yields between MC and data for
high energy ($E^\mathrm{c.m.s}_\gamma>2.8$ GeV) photon candidates found in the
endcaps of the ECL. This sample after continuum subtraction is a clean
sample of ECL clusters from beam backgrounds. 

The ratios of data and MC efficiencies versus $E^\mathrm{c.m.s}_\gamma$ are fitted using first or
second order polynomials, which are used to scale the background MC. 
Most are found to be statistically compatible with unity. An example
is the effect of the $\pi^0$ veto on photons from $\pi^0$s that escape
the veto in the partially reconstructed $D^{\ast}$ sample, which is shown in Fig.~\ref{fig:pi0vetopartialpi0}.

%%%%%%%%%%%%%%%%%%%%%%%%%%%%%%%%%%%%%%%%%%%%%%%%%%%%%%%%%%%%%%%%%%%%%%%%%%%%%%%%%
\begin{figure}
  \begin{tabular}{cc}
\hspace{-5mm}  \includegraphics[scale=0.22]{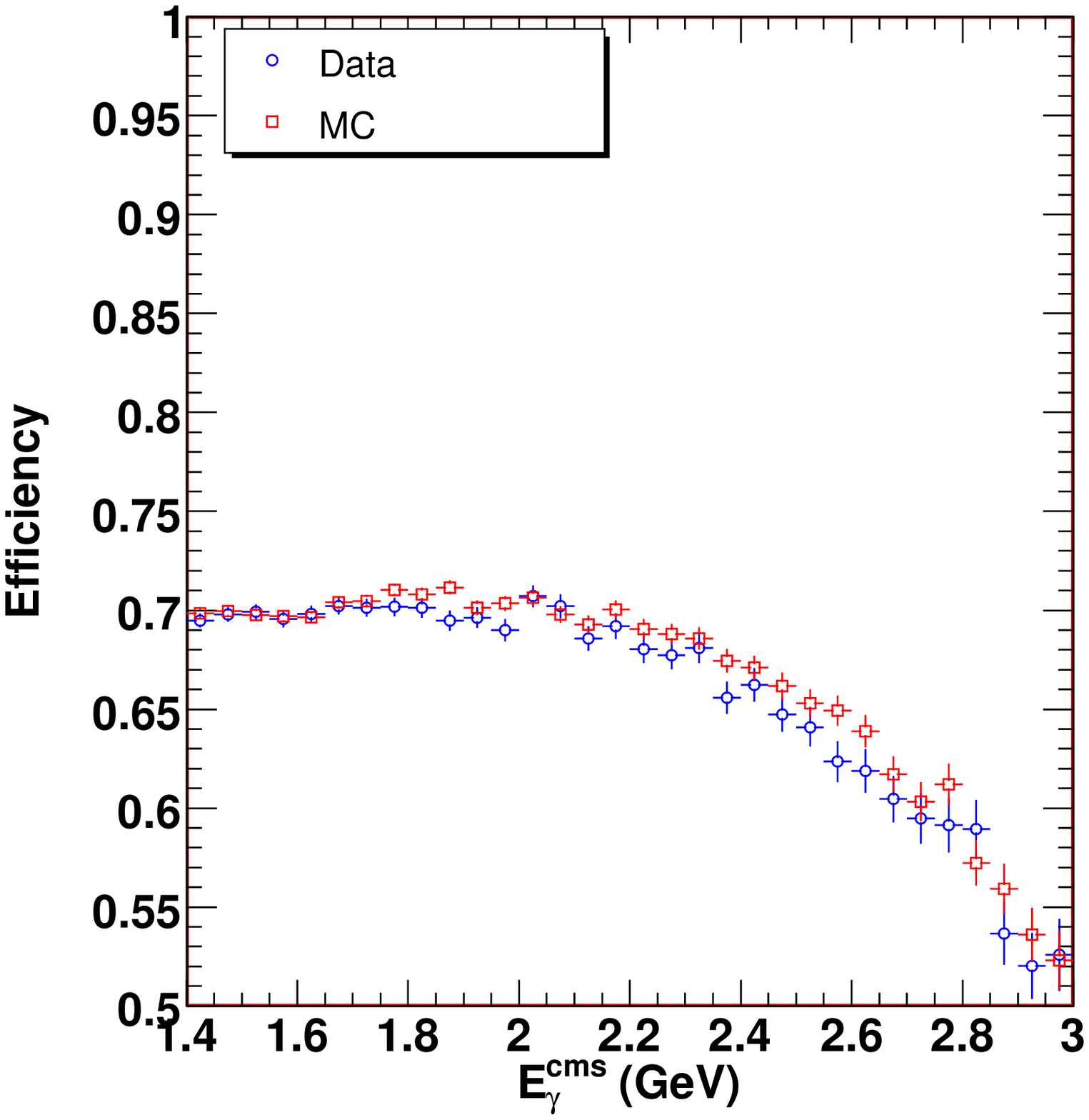} & 
\hspace{-5mm}  \includegraphics[scale=0.22]{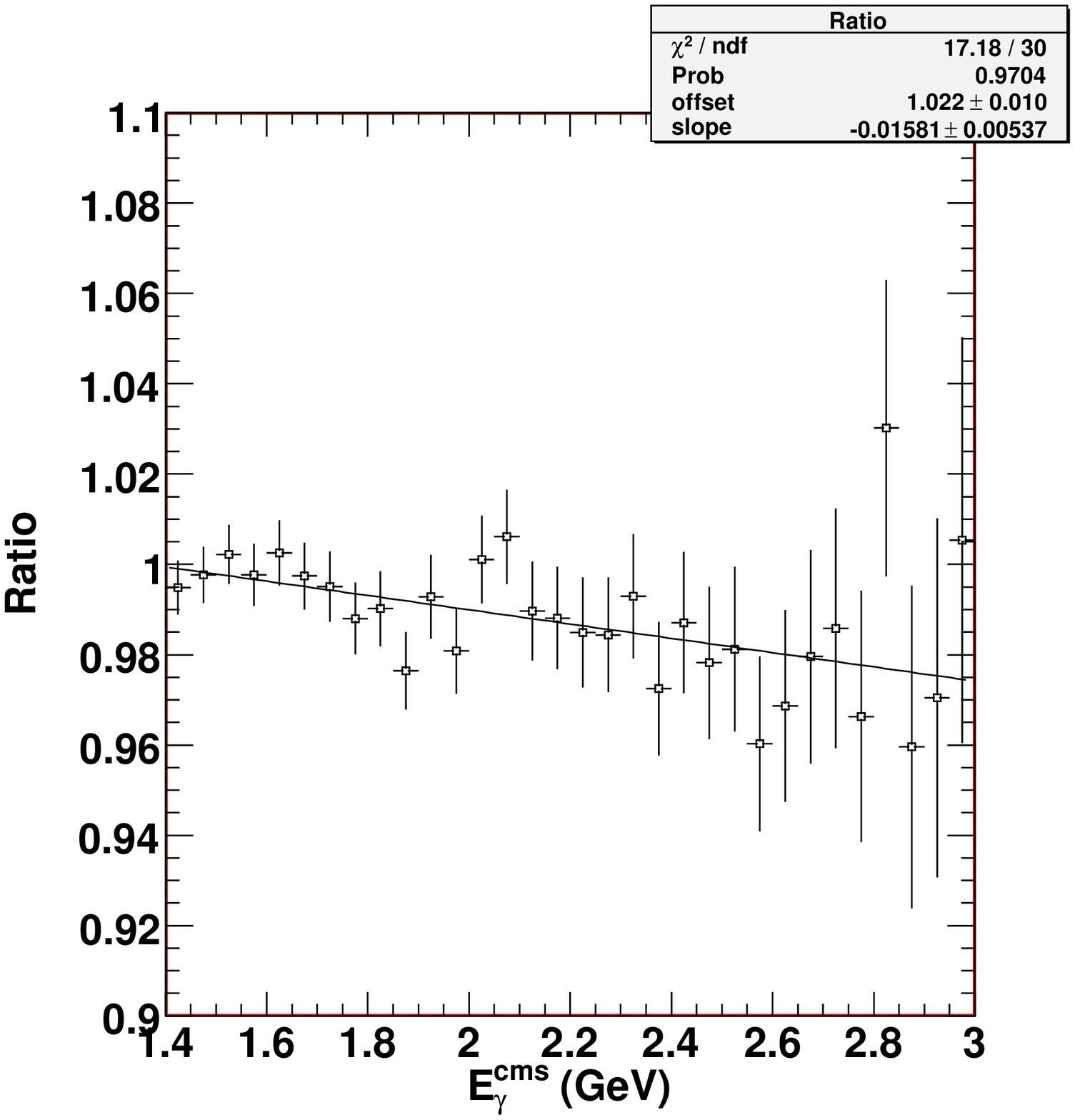} \\
\end{tabular}
    \caption{(LEFT) The $\pi^0$ veto efficiency in the partially reconstructed
      $D^{\ast}$ sample for both Data (circles) and MC (squares) and (RIGHT) their ratio
      fitted with a first order polynomial. 
      \label{fig:pi0vetopartialpi0}}	
\end{figure}
%%%%%%%%%%%%%%%%%%%%%%%%%%%%%%%%%%%%%%%%%%%%%%%%%%%%%%%%%%%%%%%%%%%%%%%%%%%%%%%%%

An exception is the efficiency of the requirement that $95$\% of the energy
be deposited in the central nine cells of the $5\times5$ cluster,
which is found to be poorly modelled by our MC for non-photon
backgrounds. 
We estimate the efficiency for data using a sample 
of candidate photons in OFF-subtracted ON data after subtracting the
known contribution from real photons.
This 
%effectively 
increases the
yield of background {\it (iv)} by 50\%.
The yield from the six background categories, after having been 
properly scaled by the above described procedures, are subtracted from 
the OFF-subtracted spectrum. The result is shown in Fig.~\ref{fig:signal}. 
After these subtractions the yield in the spectrum above the endpoint of $B$
decays is compatible with zero, $1245 \pm 4349$ candidates.
%events in the energy range 2.8 -- 4.0 GeV.

To measure the branching fraction and the moments we correct the
raw spectrum using a three step procedure: 
{\it (i)} divide by the efficiency of
the selection criteria {\it i.e.} the probability of a photon
candidate passing cuts given a
cluster has been found in the ECL, as a function of the
measured energy in the
c.m.s frame;
{\it (ii)} perform an unfolding procedure based on the Singular Value
Decomposition (SVD) algorithm~\cite{Hocker:1995kb}, which maps
the spectrum from measured energy to true
energy thereby undoing the distortion caused by the ECL;
{\it (iii)} divide by the efficiency of detection {\it i.e.} the
probability that a photon originating at the interaction point is
reconstructed in the ECL, as a function of the
true energy. Data are divided into $50$ MeV wide bins.
Step {\it (ii)}, which was not
performed in our previous analysis, is essential for a consistent extraction of
partial branching fractions and moments as a function of lower energy
thresholds. The unfolding matrix, derived from
signal MC, is calibrated to data using the results of a study of radiative di-muon
events, which gave the ECL response in data and MC in an energy
and acceptance range consistent with our analysis. We use five signal
models: KN~\cite{Kagan:1998ym}, BLNP~\cite{Lange:2005qn,Lange:2005yw}, DGE~\cite{Andersen:2006hr}, BBU~\cite{Benson:2004sg}
and GG~\cite{Gambino:2008}.
Values of the parameters of the signal model used in the signal MC
are derived from fits to the signal spectrum shown in
Fig~\ref{fig:signal}. 
The two error bars for each point show the statistical
and the total error, including the systematic error which is correlated among the points.
In order to obtain the total \bgs\ branching fraction we apply
corrections for the contribution from Cabibbo suppressed $B \to X_d\gamma$
decays.
The ratio of the \bgs\ and $B\to X_d\gamma$ branching fractions 
is assumed to be 
$R_{d/s}=(4.0\pm0.4)$\%~\cite{Hurth:2003dk}.
We apply corrections to derive the measurements in the $B$-meson rest frame, using a toy MC approach. We
generate photon 4-momentum in the rest frame of the $B$-meson
using signal models referred to earlier, and generate $B$-meson
4-momentum using their known fixed energy and
$1-\cos\theta^2$ distribution in the c.m.s.
Repeating this exercise many times yields photon energy spectra in
the rest frame of the $B$-meson and the c.m.s, from which we extract
%differences between the partial branching fraction, mean and variance
%that are the
corrections used to yield measurements in the $B$-meson
frame. The correction is derived as a mean over all signal models
while the root-mean-square is assigned as the uncertainty. 
After correcting for the acceptance we derive distributions of the partial branching
fraction, first moment (mean) and second central moment (variance) of
$B\to X_s\gamma$ as measured in the c.m.s and $B$ rest frame for lower
energy thresholds as shown in Fig.~\ref{fig:moments}. 
In the range from $1.7$ to $2.8\,\GeV$ in the rest frame of the
$B$-meson, we obtain a partial branching fraction, and the first two moments of the energy spectrum:
\begin{equation*}
  \mathcal{B}\left( B\to X_s \gamma \right) = \left( 3.31 \pm 0.19
  \pm 0.37 \pm 0.01 \right )\times10^{-4}
\end{equation*}
\begin{equation*}
  \left< E_\gamma \right> = 2.281 \pm 0.032 \pm 0.053 \pm 0.002\,\GeV
\end{equation*}
\begin{equation*}
  \left <E_\gamma^2\right>-\left<E_\gamma\right>^2 = 0.0396 \pm 0.0156 \pm 0.0214 \pm 0.0012\,\GeV^2,
\end{equation*}
where the errors are statistical, systematic and from the boost
correction, respectively.

%We correct this spectrum for the signal selection efficiency function
%obtained from signal MC, applying the same data/MC correction factors
%as for the generic photon background category {\it (iii)}.
%The average signal selection efficiency is $23$\%.

%%%%%%%%%%%%%%%%%%%%%%%%%%%%%%%%%%%%%%%%%%%%%%%%%%%%%%%%%%%%%%%%%%%%%%%%%%%%%%%%%
%\begin{figure}
%  \begin{tabular}{cc}
%\hspace{-5mm} \includegraphics[scale=0.22]{Data_e7_55} &
%\hspace{-5mm} \includegraphics[scale=0.22]{Raw_spectrum_total} \\
%\hspace{-5mm} \includegraphics[scale=0.22]{Partial_branching_fraction_total} &
%\hspace{-5mm} \includegraphics[scale=0.22]{Partial_branching_fraction_total_B_meson} \\
%\hspace{-5mm} \includegraphics[scale=0.22]{Mean_total} &
%\hspace{-5mm} \includegraphics[scale=0.22]{Mean_total_B_meson} \\
%\hspace{-5mm} \includegraphics[scale=0.22]{Variance_total} &
%\hspace{-5mm} \includegraphics[scale=0.22]{Variance_total_B_meson} \\
%\end{tabular}
%\caption{\label{fig:spectra}
%(a) ON data (open circle), scaled OFF data (open square) and
%  continuum background subtracted (filled circle) photon energy
%  spectra in the $\Upsilon(4S)$ frame.
%The two error bars show the statistical and total errors.
%}
%\end{figure}
%%%%%%%%%%%%%%%%%%%%%%%%%%%%%%%%%%%%%%%%%%%%%%%%%%%%%%%%%%%%%%%%%%%%%%%%%%%%%%%%%

%%%%%%%%%%%%%%%%%%%%%%%%%%%%%%%%%%%%%%%%%%%%%%%%%%%%%%%%%%%%%%%%%%%%%%%%%%%%%%%%%
\begin{figure}
  \begin{tabular}{c}
\hspace{-5mm} \includegraphics[scale=0.44]{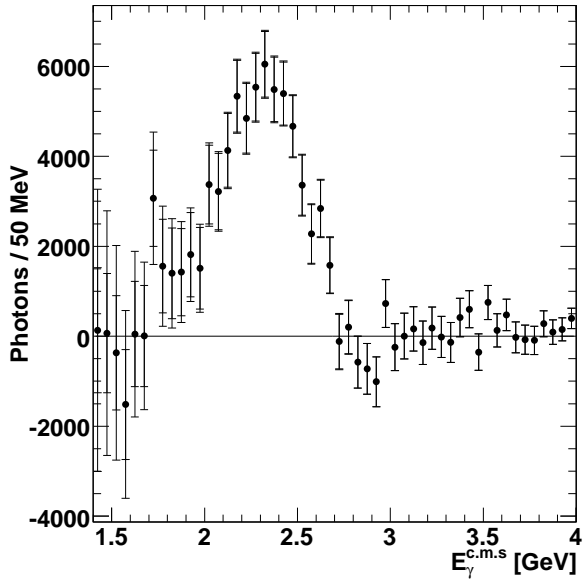} \\
\end{tabular}
  \caption{\label{fig:signal}
    The extracted photon energy spectrum of $B\to X_{s,d}\gamma$. 
    The two error bars show the statistical and total errors.
}
\end{figure}
%%%%%%%%%%%%%%%%%%%%%%%%%%%%%%%%%%%%%%%%%%%%%%%%%%%%%%%%%%%%%%%%%%%%%%%%%%%%%%%%%

%%%%%%%%%%%%%%%%%%%%%%%%%%%%%%%%%%%%%%%%%%%%%%%%%%%%%%%%%%%%%%%%%%%%%%%%%%%%%%%%%
\begin{figure}
  \begin{tabular}{cc}
\hspace{-5mm} \includegraphics[scale=0.22]{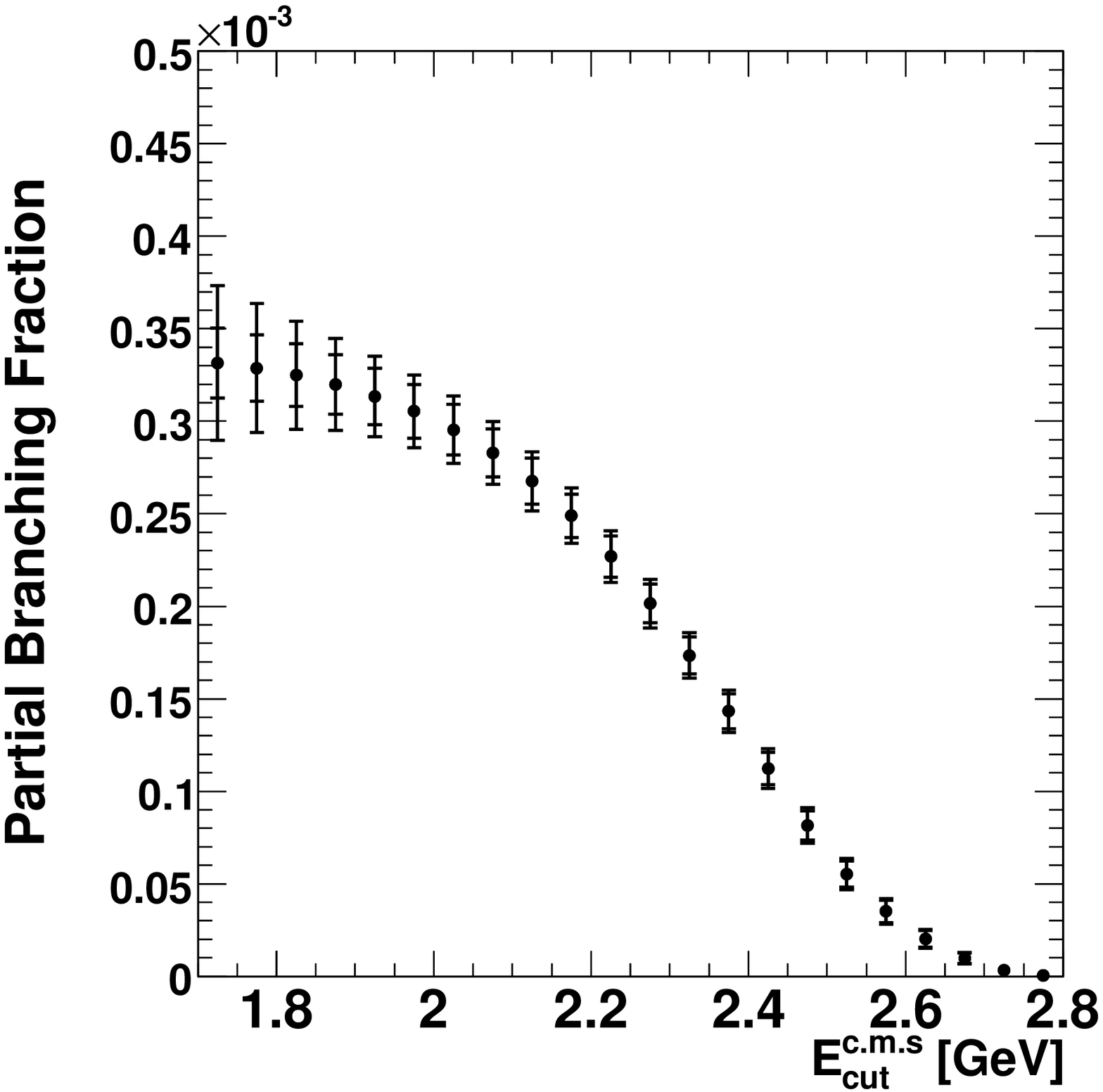} &
\hspace{-5mm} \includegraphics[scale=0.22]{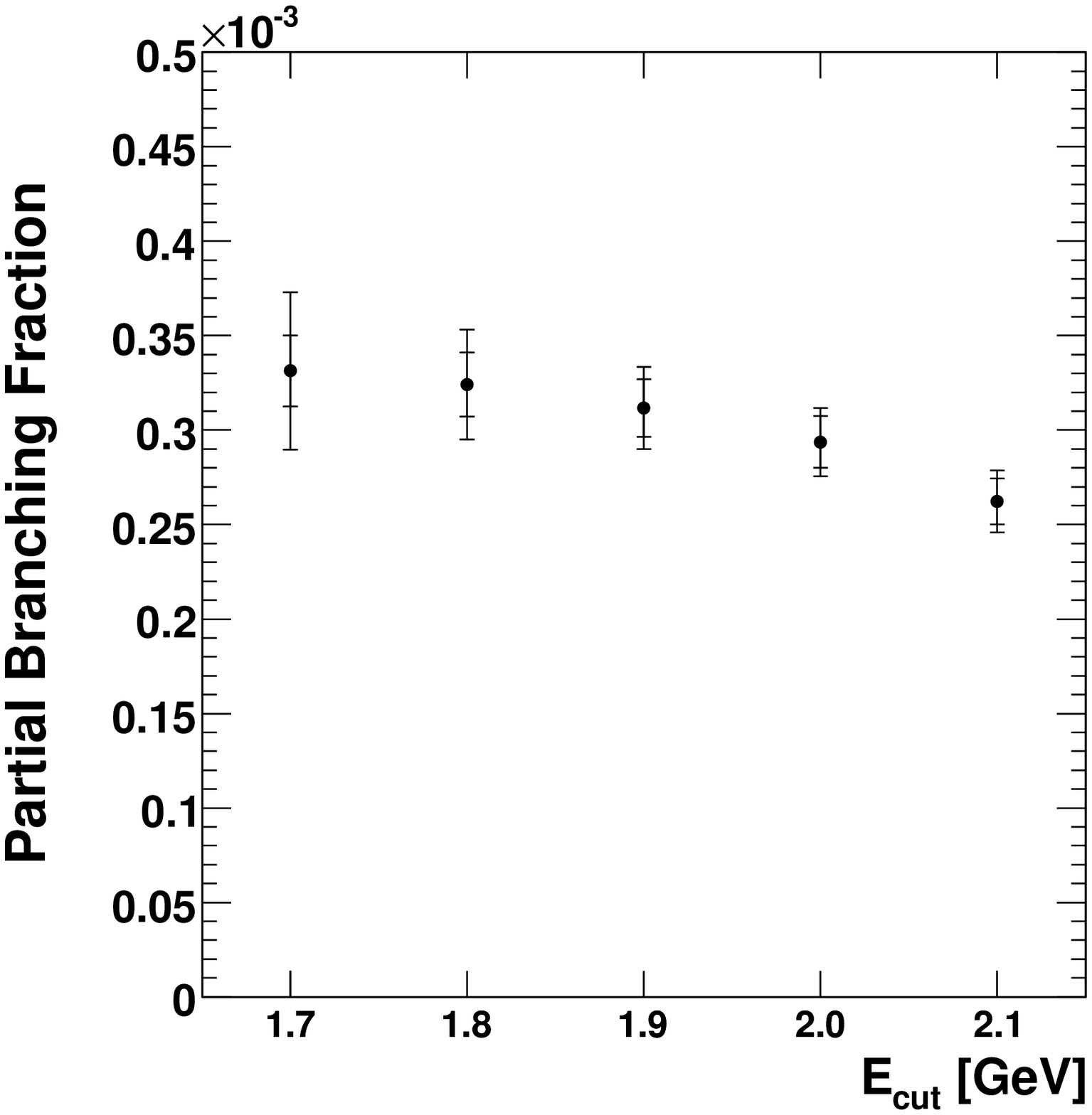} \\
\hspace{-5mm} \includegraphics[scale=0.22]{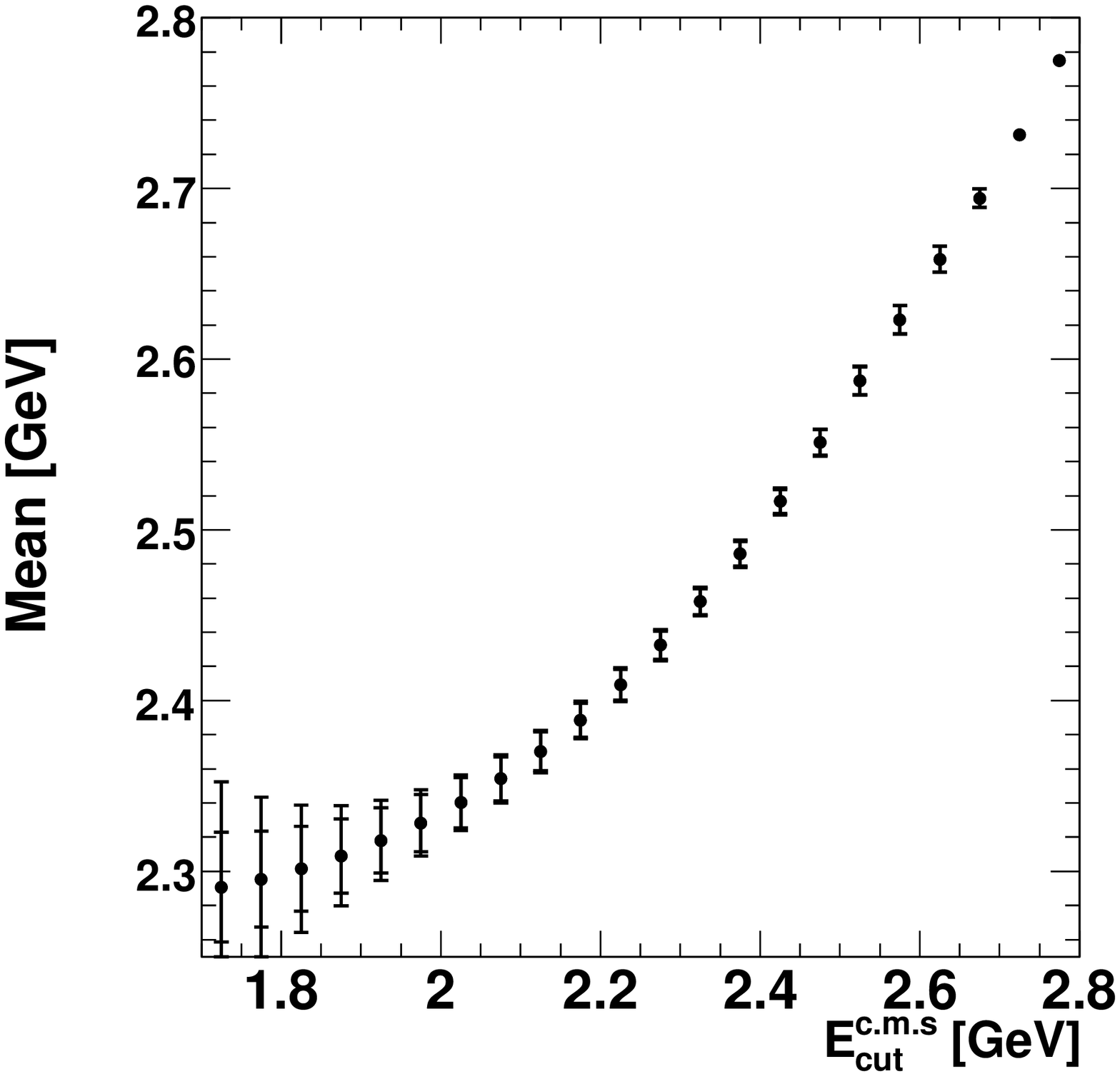} &
\hspace{-5mm} \includegraphics[scale=0.22]{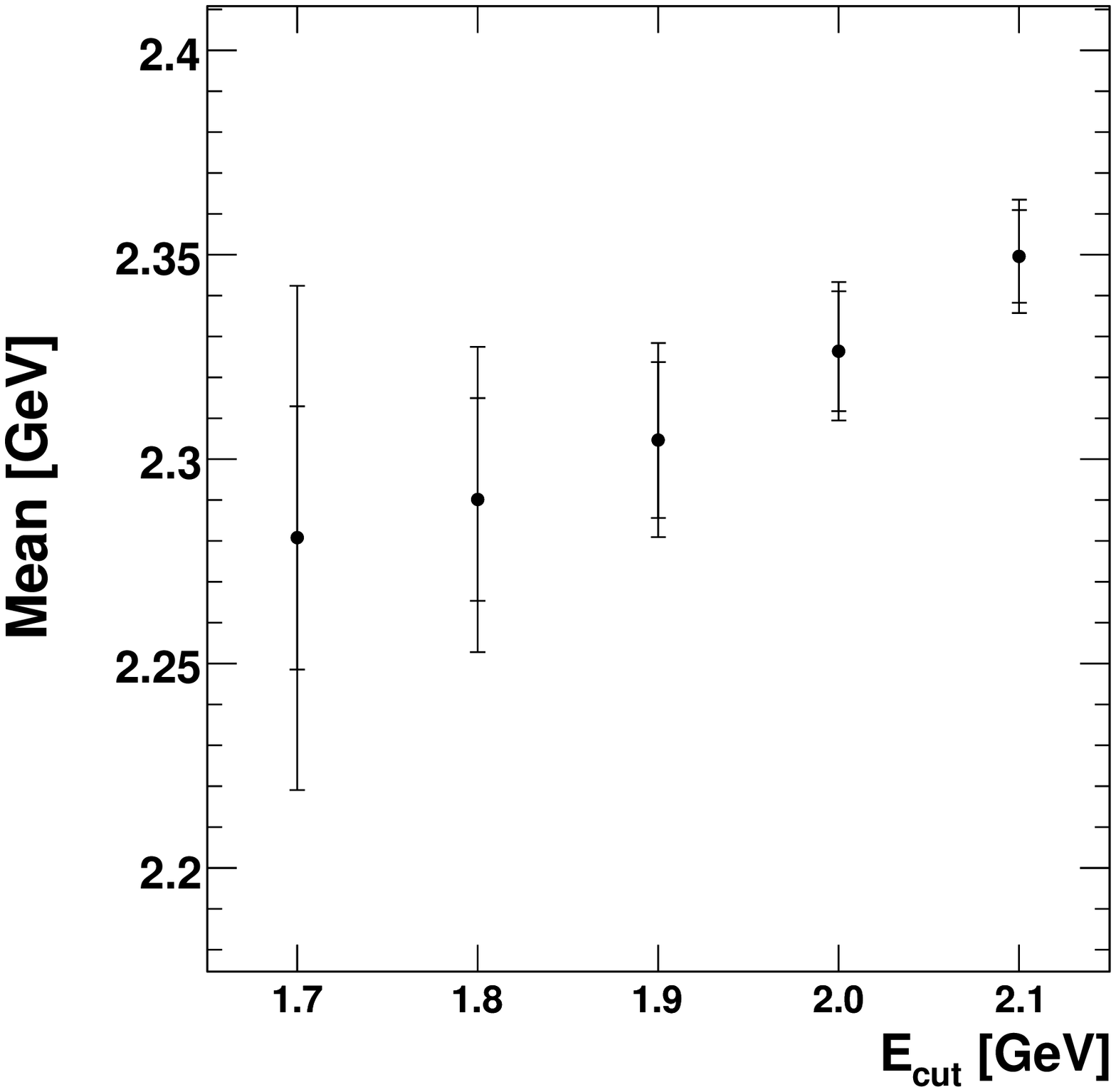} \\
\hspace{-5mm} \includegraphics[scale=0.22]{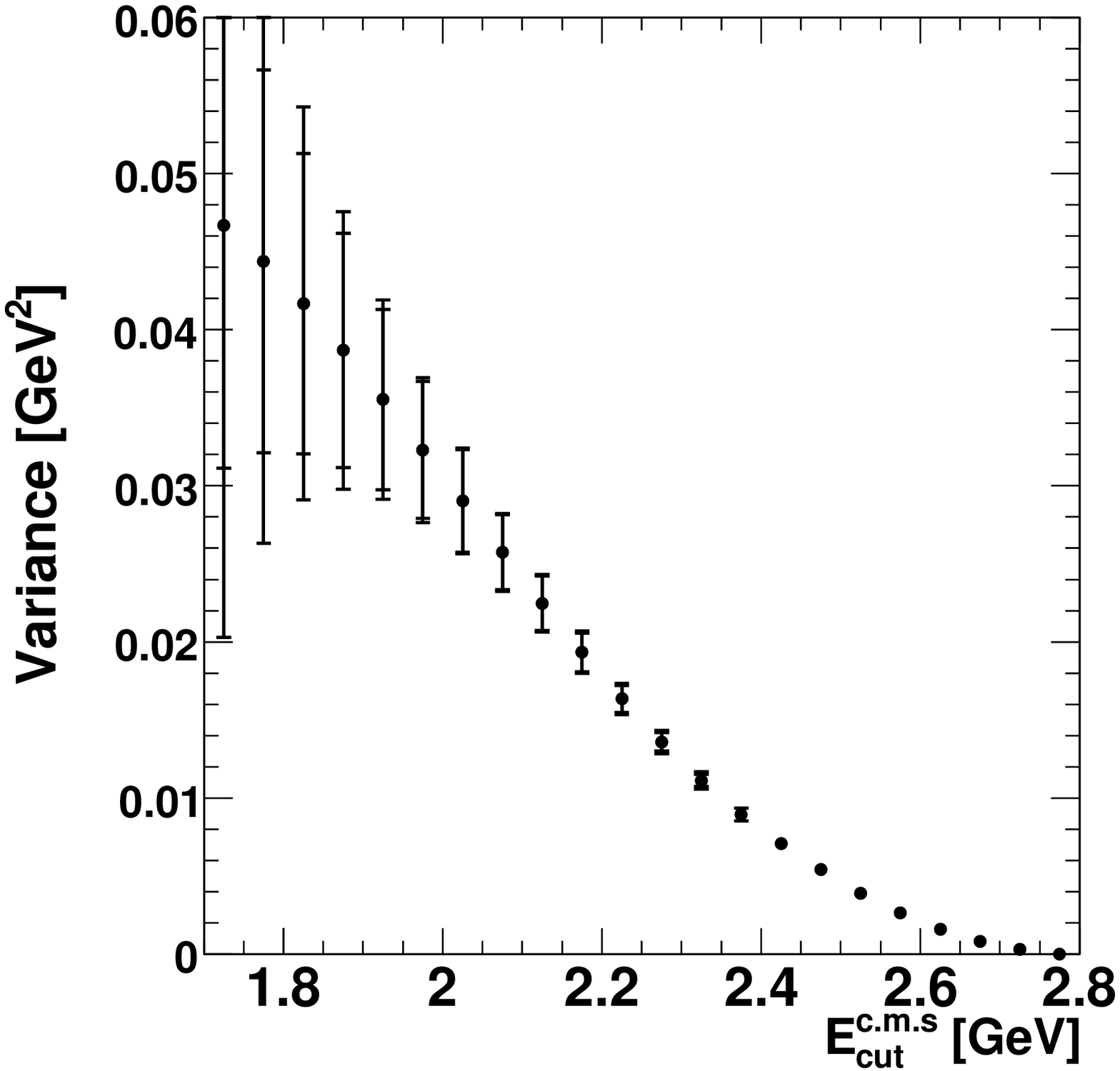} &
\hspace{-5mm} \includegraphics[scale=0.22]{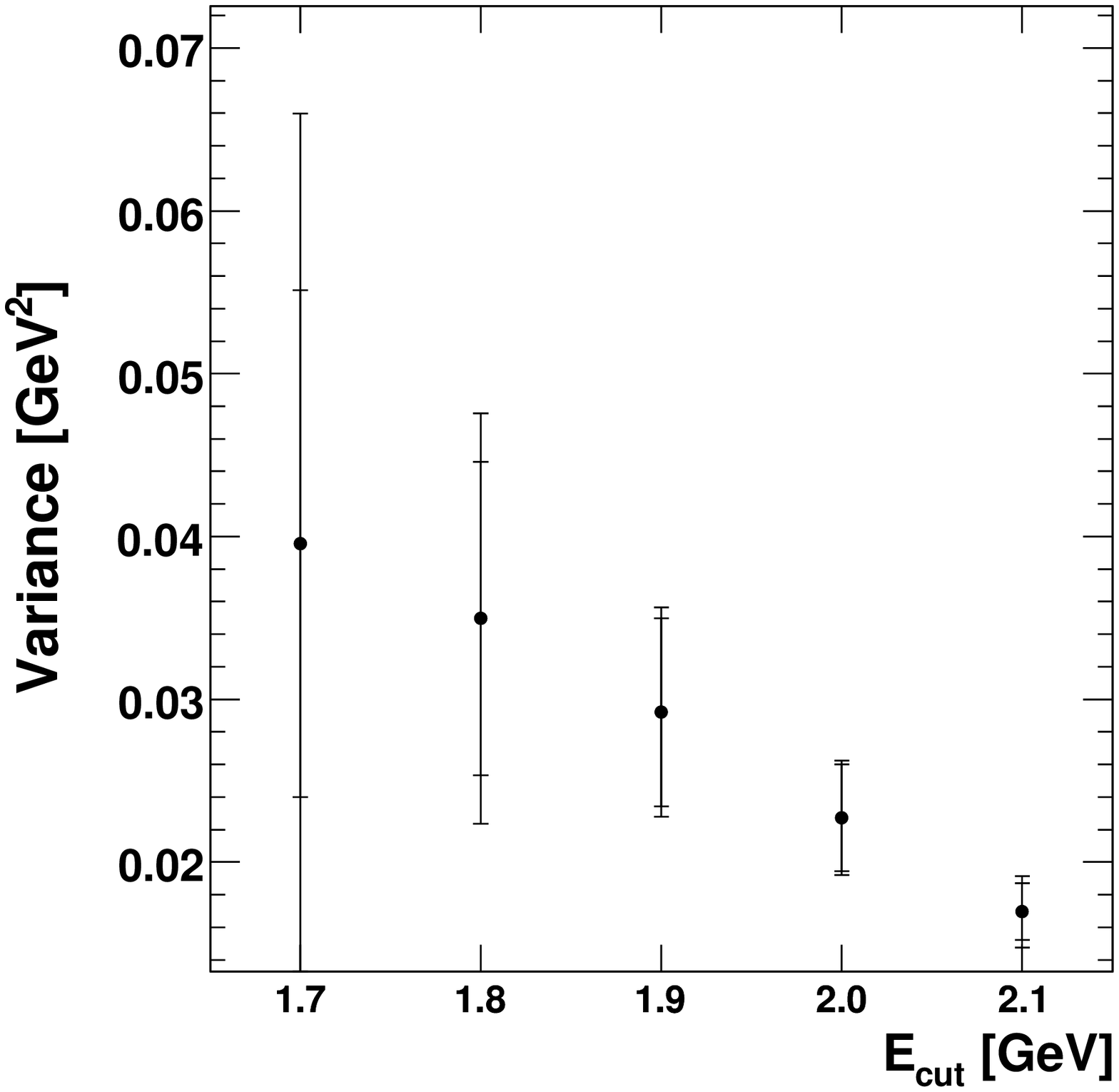} \\
\end{tabular}
  \caption{\label{fig:moments}
    (TOP) Partial branching fractions, (MIDDLE) mean, and (BOTTOM)
    variance of $B\to X_s\gamma$ in the (LEFT) c.m.s
    and (RIGHT) and in the rest frame of the $B$-meson for lower energy thresholds. 
    The two error bars show the statistical and total errors.
}
\end{figure}
%%%%%%%%%%%%%%%%%%%%%%%%%%%%%%%%%%%%%%%%%%%%%%%%%%%%%%%%%%%%%%%%%%%%%%%%%%%%%%%%%

%%%%%%%%%%%%%%%%%%%%%%%%%%%%%%%%%%%%%%%%%%%%%%%%%%
\begin{table*}
  \caption{\label{Tab} The measurements and correlation coefficients of the
    branching fraction, mean and variance of the photon energy
    spectrum for various lower energy thresolds,
    $E^B_\gamma$, as measured in the rest frame of the $B$-meson and the 
    contributions to the systematic uncertainty.
%\label{tab:table3}This is a wide table that spans the page
%width in \texttt{twocolumn} mode. It is formatted using the
%\texttt{table*} environment. It also demonstates the use of
%\textbackslash\texttt{multicolumn} in rows with entries that span
%more than one column.
  }
\begin{ruledtabular}
  \begin{tabular}{|r||rrrrr|rrrrr|rrrrr|}\tiny
    & \multicolumn{5}{c|}{$\mathcal{B}(B\to X_s\gamma)$ ($10^{-4}$)} &
    \multicolumn{5}{c|}{$\left< E_\gamma \right>$ (GeV)}  &
    \multicolumn{5}{c|}{$\Delta
    E_\gamma^2\equiv\left<E_\gamma^2\right>-\left<E_\gamma\right>^2$ (GeV$^2$)}
    \\ \hline
$E^{\mathrm{B}}_{\gamma}$ [GeV] & 1.7&    1.8&    1.9&    2.0&    2.1&    1.7 &   1.8 &   1.9 &   2.0 &   2.1 &   1.7   & 1.8   & 1.9  & 2.0   & 2.1\\\hline
Value&       3.31&    3.24&    3.12&    2.94&    2.62&    2.281&   2.290&   2.305&   2.326&   2.350&    0.0396&  0.0350&  0.0292&  0.0227&  0.0170\\
$\pm$statistical &  0.19&    0.17&    0.15&    0.14&    0.12&    0.032&   0.025&   0.019&   0.015&   0.011&    0.0156&  0.0096&  0.0058&  0.0033&  0.0017\\
$\pm$systematic &  0.37&    0.24&    0.16&    0.12&    0.10&    0.053&   0.028&   0.014&   0.007&   0.005&    0.0214&  0.0081&  0.0027&  0.0009&  0.0006\\
$\pm$boost & 0.01&    0.01&    0.02&    0.02&    0.05&    0.002&   0.002&   0.004&   0.005&   0.006&    0.0012&  0.0005&  0.0008&  0.0009&  0.0012\\\hline
\multicolumn{16}{|c|}{Systematic Uncertainties} \\\hline
Continuum  &0.18   &0.11    &0.08    &0.07    &0.07    &0.030   &0.016   &0.008   &0.004 &0.002  &0.0101  &0.0040  &0.0012  &0.0004  &0.0004\\
Selection                      &0.20   &0.15    &0.11    &0.08    &0.06    &0.023   &0.012   &0.006   &0.003 &0.001  &0.0114  &0.0039  &0.0014  &0.0005  &0.0001\\
$\pi^0/\eta$                  &0.07   &0.05    &0.04    &0.02    &0.01    &0.012   &0.006   &0.003   &0.002 &0.001  &0.0075  &0.0023  &0.0007  &0.0003  &0.0001\\
Other $B$                     &0.24   &0.13    &0.06    &0.02    &0.01    &0.033   &0.016   &0.007   &0.002 &0.000  &0.0124  &0.0051  &0.0017  &0.0004  &0.0000\\
Beam         & 0.02&    0.02&    0.01&    0.01&    0.01&    0.001   &0.001   &0.000   &0.000   &0.000    &0.0006  &0.0003  &0.0001  &0.0000  &0.0000\\
resolution       & 0.01&    0.01&    0.02&    0.02&    0.03&    0.006   &0.005   &0.005   &0.004   &0.004    &0.0009  &0.0006  &0.0005  &0.0004  &0.0004\\
Unfolding           & 0.01&    0.00&    0.00&    0.01&    0.01&    0.002   &0.001   &0.001   &0.001   &0.002    &0.0014  &0.0008  &0.0006  &0.0003  &0.0001\\
Model            & 0.03&    0.02&    0.01&    0.00&    0.00&    0.005   &0.003   &0.002   &0.001   &0.000    &0.0014  &0.0006  &0.0002  &0.0000  &0.0000\\
$\gamma$ Detection        & 0.03&    0.02&    0.01&    0.00&    0.00&    0.005   &0.003   &0.002   &0.001   &0.000    &0.0014  &0.0006  &0.0002  &0.0000  &0.0000\\
$B\to X_d\gamma$        & 0.01&    0.01&    0.01&    0.01&    0.01&
    0.000   &0.000   &0.000   &0.000   &0.000    &0.0001  &0.0000
    &0.0000  &0.0000  &0.0000\\\hline
\multicolumn{16}{|c|}{Correlation coefficients (combined statistical and
  systematic)} \\ \hline
& \multicolumn{5}{c|}{$\Delta\mathcal{B}$} & \multicolumn{5}{c|}{$\left<
  E_\gamma \right>$} & \multicolumn{5}{c|}{$\Delta E_\gamma^2$} \\
&  1.7&    1.8&    1.9&    2.0&    2.1&    1.7 &   1.8 &   1.9 &   2.0 &   2.1 &   1.7   & 1.8   & 1.9  & 2.0   & 2.1\\\hline
1.7 & 1.000 &0.959 & 0.811 & 0.699 & 0.604 &  0.455 &  0.322 & 0.114 &-0.083 &-0.142 & 0.848 &  0.857 & 0.722 & 0.528 & 0.445\\ 
1.8 &  &1.000 & 0.942 & 0.839 & 0.720 &  0.269 &  0.129 &-0.073 &-0.251 &-0.291 & 0.807 &  0.878 & 0.822 & 0.678 & 0.568\\ 
$\Delta\mathcal{B}$ 1.9 &  & & 1.000 & 0.939 & 0.823 &  0.031 & -0.107 &-0.291 &-0.442 &-0.464 & 0.680 &  0.817 & 0.869 & 0.814 & 0.700\\ 
2.0 &  & &  & 1.000 & 0.959 & -0.004 & -0.143 &-0.332 &-0.494 &-0.531 & 0.612 &  0.767 & 0.863 & 0.870 & 0.846\\ 
2.1 &  & &  &  & 1.000 &  0.023 & -0.107 &-0.296 &-0.476 &-0.548 & 0.546 &  0.689 & 0.795 & 0.848 & 0.910\\ 
1.7 &       &      &       &       &       &  1.000 &  0.967 & 0.838 & 0.636 & 0.489 &  0.342 &  0.149 &-0.094 &-0.252 &-0.174\\
1.8 &       &      &       &       &       &   &  1.000 & 0.946 & 0.793 & 0.645 &  0.155 & -0.047 &-0.290 &-0.431 &-0.329\\
$\left< E_\gamma \right>$ 1.9&       &      &       &       &       &   &   & 1.000 & 0.942 & 0.824 & -0.066 & -0.280 &-0.516 &-0.640 &-0.530\\
2.0&       &      &       &       &       &   &   &  & 1.000 & 0.954 & -0.230 & -0.438 &-0.660 &-0.779 &-0.696\\
2.1&       &      &       &       &       &   &   &  & & 1.000 & -0.252 & -0.438 &-0.642 &-0.777 &-0.751\\
1.7&       &      &       &       &       &        &        &       &       &       &  1.000 &  0.945 & 0.782 & 0.581 & 0.497\\ 
1.8&       &      &       &       &       &        &        &       &       &       &   &  1.000 & 0.935 & 0.782 & 0.677\\ 
$\Delta E_\gamma^2$ 1.9&       &      &       &       &       &        &        &       &       &       &   &   & 1.000 & 0.946 & 0.840\\ 
2.0&       &      &       &       &       &        &        &       &       &       &   &   & & 1.000 & 0.942\\ 
2.1&       &      &       &       &       &        &        &       &       &       &   &   & & & 1.000\\ 
 \end{tabular}
  \end{ruledtabular}    
\end{table*}
\normalsize
%%%%%%%%%%%%%%%%%%%%%%%%%%%%%%%%%%%%%%%%%%%%%%%%%%
The full results, the systematic error budget and correlation
coefficients for five lower energy thresholds
($E^B_\gamma=1.7,\,1.8,\,1.9,\,2.0,\,2.1$ GeV) are listed in Table~\ref{Tab}.
The total systematic error is derived from a sum in quadrature over
all sources. We vary the number of $B\bar{B}$, the ON to OFF ratio of integrated
luminosities and the correction factors applied to the OFF data photon
candidates and assign the observed variation as the systematic
associated with continuum subtraction. 
The parameters of the correction functions applied to the $\pi^0$ and
$\eta$ yields are varied taking into account their correlations. 
As we do not measure the yields of photons from sources other than 
    $\pi^0$'s and $\eta$'s in $B\bar{B}$ events, we independently vary the 
    expected yields of these additional sources by $\pm20$\%.
For the model dependence in correcting for the acceptance we use four
signal models in addition to the default model, and assign the maximum
deviation from the default as the uncertainty. 
The error on the photon detection efficiency in the ECL is
measured to be $2$\% using radiative $\mu$-pair events, and 
also affects the estimation of photons from
    $B\bar{B}$. For the uncertainties related to the unfolding
procedure, we vary the effective rank parameter up and down by one in
the SVD algorithm.
% From Tim
%using recent values for the CKM matrix elements~\cite{CKMWS},
%taking into account errors due to the
%choice of the charm mass~\cite{Gambino:2001ew} and a $10$\%
%error for the unknown $u\bar{u}$-loop contributions.
%  
%  which is obtained using recent values for the CKM matrix elements~\cite{CKMWS},
%  uncertainty of the charm quark mass~\cite{Gambino:2001ew}
%  and a $10$\% error for the unknown $u\bar{u}$-loop contributions.
%The selection efficiency for $b\to d\gamma$ is found to be equal
%to the efficiency for \bgs\ within 10\%, which we include in the
%systematic error.

In conclusion, we have measured the branching fraction 
and photon energy spectrum of \bgs\ in the 
energy range $1.7\,\GeV\le E^\mathrm{c.m.s}_\gamma\le2.8\,\GeV$ in a fully 
inclusive way.
For the first time 97\% of the spectrum is
measured~\cite{Buchmuller:2005zv} allowing the theoretical
uncertainties to be reduced to a very low level. 
Using $605\ifb$ of data taken at the $\Upsilon(4S)$ and
$68\ifb$ taken below the resonance, we obtain
${\mathcal B}(B\to X_s\gamma : E^B_\gamma>1.7\,\mathrm{GeV})= \left(3.31 \pm 0.19 \pm 0.37 \pm 0.01\right)\times 10^{-4}$,
where the errors are statistical, systematic and due to the boost correction, respectively. 
This result is in agreement with the latest theoretical 
calculations~\cite{Misiak:2006zs,Becher:2006pu,Andersen:2006hr}.
The results can be used to place constraints on new
physics~\cite{Buchmueller:2007zk}
and determine SM parameters such as the $b$-quark mass~\cite{Schwanda:2008kw}.

\begin{acknowledgments}
We thank the KEKB group for excellent operation of the
accelerator, the KEK cryogenics group for efficient solenoid
operations, and the KEK computer group and
the NII for valuable computing and SINET3 network
support.  We acknowledge support from MEXT and JSPS (Japan);
ARC, DEST and A.J. Slocum (Australia); NSFC (China); 
DST (India); MOEHRD, KOSEF and KRF (Korea); 
KBN (Poland); MES and RFAAE (Russia); ARRS (Slovenia); SNSF (Switzerland); 
NSC and MOE (Taiwan); and DOE (USA).
\end{acknowledgments}

\bibliography{manuscript}% Produces the bibliography via BibTeX.

\begin{thebibliography}{33}
\expandafter\ifx\csname natexlab\endcsname\relax\def\natexlab#1{#1}\fi
\expandafter\ifx\csname bibnamefont\endcsname\relax
  \def\bibnamefont#1{#1}\fi
\expandafter\ifx\csname bibfnamefont\endcsname\relax
  \def\bibfnamefont#1{#1}\fi
\expandafter\ifx\csname citenamefont\endcsname\relax
  \def\citenamefont#1{#1}\fi
\expandafter\ifx\csname url\endcsname\relax
  \def\url#1{\texttt{#1}}\fi
\expandafter\ifx\csname urlprefix\endcsname\relax\def\urlprefix{URL }\fi
\providecommand{\bibinfo}[2]{#2}
\providecommand{\eprint}[2][]{\url{#2}}

\bibitem[{\citenamefont{Misiak et~al.}(2007)}]{Misiak:2006zs}
\bibinfo{author}{\bibfnamefont{M.}~\bibnamefont{Misiak}} \bibnamefont{et~al.},
  \bibinfo{journal}{Phys. Rev. Lett.} \textbf{\bibinfo{volume}{98}},
  \bibinfo{pages}{022002} (\bibinfo{year}{2007}), \eprint{hep-ph/0609232}.

\bibitem[{\citenamefont{Becher and Neubert}(2007)}]{Becher:2006pu}
\bibinfo{author}{\bibfnamefont{T.}~\bibnamefont{Becher}} \bibnamefont{and}
  \bibinfo{author}{\bibfnamefont{M.}~\bibnamefont{Neubert}},
  \bibinfo{journal}{Phys. Rev. Lett.} \textbf{\bibinfo{volume}{98}},
  \bibinfo{pages}{022003} (\bibinfo{year}{2007}), \eprint{hep-ph/0610067}.

\bibitem[{\citenamefont{Yao et~al.}(2006)}]{Yao:2006px}
\bibinfo{author}{\bibfnamefont{W.~M.} \bibnamefont{Yao}} \bibnamefont{et~al.}
  (\bibinfo{collaboration}{Particle Data Group}), \bibinfo{journal}{J. Phys.}
  \textbf{\bibinfo{volume}{G33}}, \bibinfo{pages}{1} (\bibinfo{year}{2006}).

\bibitem[{\citenamefont{Bertolini et~al.}(1991)\citenamefont{Bertolini,
  Borzumati, Masiero, and Ridolfi}}]{Bertolini:1990if}
\bibinfo{author}{\bibfnamefont{S.}~\bibnamefont{Bertolini}},
  \bibinfo{author}{\bibfnamefont{F.}~\bibnamefont{Borzumati}},
  \bibinfo{author}{\bibfnamefont{A.}~\bibnamefont{Masiero}}, \bibnamefont{and}
  \bibinfo{author}{\bibfnamefont{G.}~\bibnamefont{Ridolfi}},
  \bibinfo{journal}{Nucl. Phys.} \textbf{\bibinfo{volume}{B353}},
  \bibinfo{pages}{591} (\bibinfo{year}{1991}).

\bibitem[{\citenamefont{Cho and Misiak}(1994)}]{Cho:1993zb}
\bibinfo{author}{\bibfnamefont{P.~L.} \bibnamefont{Cho}} \bibnamefont{and}
  \bibinfo{author}{\bibfnamefont{M.}~\bibnamefont{Misiak}},
  \bibinfo{journal}{Phys. Rev.} \textbf{\bibinfo{volume}{D49}},
  \bibinfo{pages}{5894} (\bibinfo{year}{1994}), \eprint{hep-ph/9310332}.

\bibitem[{\citenamefont{Fujikawa and Yamada}(1994)}]{Fujikawa:1993zu}
\bibinfo{author}{\bibfnamefont{K.}~\bibnamefont{Fujikawa}} \bibnamefont{and}
  \bibinfo{author}{\bibfnamefont{A.}~\bibnamefont{Yamada}},
  \bibinfo{journal}{Phys. Rev.} \textbf{\bibinfo{volume}{D49}},
  \bibinfo{pages}{5890} (\bibinfo{year}{1994}).

\bibitem[{\citenamefont{Barberio et~al.}(2007)}]{Barberio:2007cr}
\bibinfo{author}{\bibfnamefont{E.}~\bibnamefont{Barberio}} \bibnamefont{et~al.}
  (\bibinfo{collaboration}{Heavy Flavor Averaging Group (HFAG)})
  (\bibinfo{year}{2007}), \eprint{arXiv:0704.3575 [hep-ex]}.

\bibitem[{\citenamefont{Bigi and Uraltsev}(2002)}]{Bigi:2002qq}
\bibinfo{author}{\bibfnamefont{I.}~\bibnamefont{Bigi}} \bibnamefont{and}
  \bibinfo{author}{\bibfnamefont{N.}~\bibnamefont{Uraltsev}},
  \bibinfo{journal}{Int. J. Mod. Phys.} \textbf{\bibinfo{volume}{A17}},
  \bibinfo{pages}{4709} (\bibinfo{year}{2002}), \eprint{hep-ph/0202175}.

\bibitem[{\citenamefont{Abe et~al.}(2001)}]{Abe:2001hk}
\bibinfo{author}{\bibfnamefont{K.}~\bibnamefont{Abe}} \bibnamefont{et~al.}
  (\bibinfo{collaboration}{Belle}), \bibinfo{journal}{Phys. Lett.}
  \textbf{\bibinfo{volume}{B511}}, \bibinfo{pages}{151} (\bibinfo{year}{2001}),
  \eprint{hep-ex/0103042}.

\bibitem[{\citenamefont{Koppenburg et~al.}(2004)}]{Koppenburg:2004fz}
\bibinfo{author}{\bibfnamefont{P.}~\bibnamefont{Koppenburg}}
  \bibnamefont{et~al.} (\bibinfo{collaboration}{Belle}) (\bibinfo{year}{2004}),
  \eprint{hep-ex/0403004}.

\bibitem[{\citenamefont{Chen et~al.}(2001)}]{Chen:2001fja}
\bibinfo{author}{\bibfnamefont{S.}~\bibnamefont{Chen}} \bibnamefont{et~al.}
  (\bibinfo{collaboration}{CLEO}), \bibinfo{journal}{Phys. Rev. Lett.}
  \textbf{\bibinfo{volume}{87}}, \bibinfo{pages}{251807}
  (\bibinfo{year}{2001}), \eprint{hep-ex/0108032}.

\bibitem[{\citenamefont{Aubert et~al.}(2005)}]{Aubert:2005cua}
\bibinfo{author}{\bibfnamefont{B.}~\bibnamefont{Aubert}} \bibnamefont{et~al.}
  (\bibinfo{collaboration}{BABAR}), \bibinfo{journal}{Phys. Rev.}
  \textbf{\bibinfo{volume}{D72}}, \bibinfo{pages}{052004}
  (\bibinfo{year}{2005}), \eprint{hep-ex/0508004}.

\bibitem[{\citenamefont{Aubert et~al.}(2006)}]{Aubert:2006gg}
\bibinfo{author}{\bibfnamefont{B.}~\bibnamefont{Aubert}} \bibnamefont{et~al.}
  (\bibinfo{collaboration}{BaBar}), \bibinfo{journal}{Phys. Rev. Lett.}
  \textbf{\bibinfo{volume}{97}}, \bibinfo{pages}{171803}
  (\bibinfo{year}{2006}), \eprint{hep-ex/0607071}.

\bibitem[{\citenamefont{Aubert et~al.}(2008)}]{Aubert:2007my}
\bibinfo{author}{\bibfnamefont{B.}~\bibnamefont{Aubert}} \bibnamefont{et~al.}
  (\bibinfo{collaboration}{BABAR}), \bibinfo{journal}{Phys. Rev.}
  \textbf{\bibinfo{volume}{D77}}, \bibinfo{pages}{051103(R)}
  (\bibinfo{year}{2008}), \eprint{arXiv:0711.4889 [hep-ex]}.

\bibitem[{\citenamefont{Kurokawa and Kikutani}(2003)}]{KEKB}
\bibinfo{author}{\bibfnamefont{S.}~\bibnamefont{Kurokawa}} \bibnamefont{and}
  \bibinfo{author}{\bibfnamefont{E.}~\bibnamefont{Kikutani}},
  \bibinfo{journal}{Nucl. Instrum. Meth.} \textbf{\bibinfo{volume}{A499}},
  \bibinfo{pages}{1} (\bibinfo{year}{2003}), \bibinfo{note}{and other papers
  included in this Volume.}

\bibitem[{\citenamefont{Abashian et~al.}(2002)}]{NIM:Belle}
\bibinfo{author}{\bibfnamefont{A.}~\bibnamefont{Abashian}}
  \bibnamefont{et~al.}, \bibinfo{journal}{Nucl. Instrum. Meth.}
  \textbf{\bibinfo{volume}{A479}}, \bibinfo{pages}{117} (\bibinfo{year}{2002}).

\bibitem[{\citenamefont{Fox and Wolfram}(1978)}]{Fox:1978vu}
\bibinfo{author}{\bibfnamefont{G.~C.} \bibnamefont{Fox}} \bibnamefont{and}
  \bibinfo{author}{\bibfnamefont{S.}~\bibnamefont{Wolfram}},
  \bibinfo{journal}{Phys. Rev. Lett.} \textbf{\bibinfo{volume}{41}},
  \bibinfo{pages}{1581} (\bibinfo{year}{1978}).

\bibitem[{\citenamefont{Brun et~al.}(1978)\citenamefont{Brun, Hagelberg,
  Hansroul, and Lassalle}}]{Brun:1978fy}
\bibinfo{author}{\bibfnamefont{R.}~\bibnamefont{Brun}},
  \bibinfo{author}{\bibfnamefont{R.}~\bibnamefont{Hagelberg}},
  \bibinfo{author}{\bibfnamefont{M.}~\bibnamefont{Hansroul}}, \bibnamefont{and}
  \bibinfo{author}{\bibfnamefont{J.~C.} \bibnamefont{Lassalle}}
  (\bibinfo{year}{1978}), \bibinfo{note}{cERN-DD-78-2-REV}.

\bibitem[{\citenamefont{Andersen and Gardi}(2005)}]{Andersen:2005bj}
\bibinfo{author}{\bibfnamefont{J.~R.} \bibnamefont{Andersen}} \bibnamefont{and}
  \bibinfo{author}{\bibfnamefont{E.}~\bibnamefont{Gardi}},
  \bibinfo{journal}{JHEP} \textbf{\bibinfo{volume}{06}}, \bibinfo{pages}{030}
  (\bibinfo{year}{2005}), \eprint{hep-ph/0502159}.

\bibitem[{\citenamefont{Casey}(2001)}]{Casey}
\bibinfo{author}{\bibfnamefont{B.}~\bibnamefont{Casey}}, Ph.D. thesis,
  \bibinfo{school}{University of Hawaii} (\bibinfo{year}{2001}),
  \bibinfo{note}{http://www.phys.hawaii.edu/~casey/thesis/}.

\bibitem[{\citenamefont{Barberio and Was}(1994)}]{Barberio:1993qi}
\bibinfo{author}{\bibfnamefont{E.}~\bibnamefont{Barberio}} \bibnamefont{and}
  \bibinfo{author}{\bibfnamefont{Z.}~\bibnamefont{Was}},
  \bibinfo{journal}{Comput. Phys. Commun.} \textbf{\bibinfo{volume}{79}},
  \bibinfo{pages}{291} (\bibinfo{year}{1994}).

\bibitem[{\citenamefont{Lange}(2001)}]{Lange:2001uf}
\bibinfo{author}{\bibfnamefont{D.~J.} \bibnamefont{Lange}},
  \bibinfo{journal}{Nucl. Instrum. Meth.} \textbf{\bibinfo{volume}{A462}},
  \bibinfo{pages}{152} (\bibinfo{year}{2001}).

\bibitem[{\citenamefont{Hocker and Kartvelishvili}(1996)}]{Hocker:1995kb}
\bibinfo{author}{\bibfnamefont{A.}~\bibnamefont{Hocker}} \bibnamefont{and}
  \bibinfo{author}{\bibfnamefont{V.}~\bibnamefont{Kartvelishvili}},
  \bibinfo{journal}{Nucl. Instrum. Meth.} \textbf{\bibinfo{volume}{A372}},
  \bibinfo{pages}{469} (\bibinfo{year}{1996}), \eprint{hep-ph/9509307}.

\bibitem[{\citenamefont{Kagan and Neubert}(1999)}]{Kagan:1998ym}
\bibinfo{author}{\bibfnamefont{A.~L.} \bibnamefont{Kagan}} \bibnamefont{and}
  \bibinfo{author}{\bibfnamefont{M.}~\bibnamefont{Neubert}},
  \bibinfo{journal}{Eur. Phys. J.} \textbf{\bibinfo{volume}{C7}},
  \bibinfo{pages}{5} (\bibinfo{year}{1999}), \eprint{hep-ph/9805303}.

\bibitem[{\citenamefont{Lange et~al.}(2005{\natexlab{a}})\citenamefont{Lange,
  Neubert, and Paz}}]{Lange:2005qn}
\bibinfo{author}{\bibfnamefont{B.~O.} \bibnamefont{Lange}},
  \bibinfo{author}{\bibfnamefont{M.}~\bibnamefont{Neubert}}, \bibnamefont{and}
  \bibinfo{author}{\bibfnamefont{G.}~\bibnamefont{Paz}},
  \bibinfo{journal}{JHEP} \textbf{\bibinfo{volume}{10}}, \bibinfo{pages}{084}
  (\bibinfo{year}{2005}{\natexlab{a}}), \eprint{hep-ph/0508178}.

\bibitem[{\citenamefont{Lange et~al.}(2005{\natexlab{b}})\citenamefont{Lange,
  Neubert, and Paz}}]{Lange:2005yw}
\bibinfo{author}{\bibfnamefont{B.~O.} \bibnamefont{Lange}},
  \bibinfo{author}{\bibfnamefont{M.}~\bibnamefont{Neubert}}, \bibnamefont{and}
  \bibinfo{author}{\bibfnamefont{G.}~\bibnamefont{Paz}},
  \bibinfo{journal}{Phys. Rev.} \textbf{\bibinfo{volume}{D72}},
  \bibinfo{pages}{073006} (\bibinfo{year}{2005}{\natexlab{b}}),
  \eprint{hep-ph/0504071}.

\bibitem[{\citenamefont{Andersen and Gardi}(2007)}]{Andersen:2006hr}
\bibinfo{author}{\bibfnamefont{J.~R.} \bibnamefont{Andersen}} \bibnamefont{and}
  \bibinfo{author}{\bibfnamefont{E.}~\bibnamefont{Gardi}},
  \bibinfo{journal}{JHEP} \textbf{\bibinfo{volume}{01}}, \bibinfo{pages}{029}
  (\bibinfo{year}{2007}), \eprint{hep-ph/0609250}.

\bibitem[{\citenamefont{Benson et~al.}(2005)\citenamefont{Benson, Bigi, and
  Uraltsev}}]{Benson:2004sg}
\bibinfo{author}{\bibfnamefont{D.}~\bibnamefont{Benson}},
  \bibinfo{author}{\bibfnamefont{I.~I.} \bibnamefont{Bigi}}, \bibnamefont{and}
  \bibinfo{author}{\bibfnamefont{N.}~\bibnamefont{Uraltsev}},
  \bibinfo{journal}{Nucl. Phys.} \textbf{\bibinfo{volume}{B710}},
  \bibinfo{pages}{371} (\bibinfo{year}{2005}), \eprint{hep-ph/0410080}.

\bibitem[{\citenamefont{Gambino and Giordano}(2008)}]{Gambino:2008}
\bibinfo{author}{\bibfnamefont{P.}~\bibnamefont{Gambino}} \bibnamefont{and}
  \bibinfo{author}{\bibfnamefont{P.}~\bibnamefont{Giordano}}
  (\bibinfo{year}{2008}), \bibinfo{note}{work in progress}.

\bibitem[{\citenamefont{Hurth et~al.}(2005)\citenamefont{Hurth, Lunghi, and
  Porod}}]{Hurth:2003dk}
\bibinfo{author}{\bibfnamefont{T.}~\bibnamefont{Hurth}},
  \bibinfo{author}{\bibfnamefont{E.}~\bibnamefont{Lunghi}}, \bibnamefont{and}
  \bibinfo{author}{\bibfnamefont{W.}~\bibnamefont{Porod}},
  \bibinfo{journal}{Nucl. Phys.} \textbf{\bibinfo{volume}{B704}},
  \bibinfo{pages}{56} (\bibinfo{year}{2005}), \eprint{hep-ph/0312260}.

\bibitem[{\citenamefont{Buchmuller and Flacher}(2006)}]{Buchmuller:2005zv}
\bibinfo{author}{\bibfnamefont{O.}~\bibnamefont{Buchmuller}} \bibnamefont{and}
  \bibinfo{author}{\bibfnamefont{H.}~\bibnamefont{Flacher}},
  \bibinfo{journal}{Phys. Rev.} \textbf{\bibinfo{volume}{D73}},
  \bibinfo{pages}{073008} (\bibinfo{year}{2006}), \eprint{hep-ph/0507253}.

\bibitem[{\citenamefont{Buchmueller et~al.}(2007)}]{Buchmueller:2007zk}
\bibinfo{author}{\bibfnamefont{O.}~\bibnamefont{Buchmueller}}
  \bibnamefont{et~al.}, \bibinfo{journal}{Phys. Lett.}
  \textbf{\bibinfo{volume}{B657}}, \bibinfo{pages}{87} (\bibinfo{year}{2007}),
  \eprint{0707.3447}.

\bibitem[{\citenamefont{Schwanda et~al.}(2008)\citenamefont{Schwanda, Urquijo,
  Barberio, Limosani et~al.}}]{Schwanda:2008kw}
\bibinfo{author}{\bibfnamefont{C.}~\bibnamefont{Schwanda}},
  \bibinfo{author}{\bibfnamefont{P.}~\bibnamefont{Urquijo}},
  \bibinfo{author}{\bibfnamefont{E.}~\bibnamefont{Barberio}},
  \bibinfo{author}{\bibfnamefont{A.}~\bibnamefont{Limosani}},
  \bibnamefont{et~al.} (\bibinfo{collaboration}{Belle}),
  \bibinfo{journal}{submitted to Phys. Rev. D.}  (\bibinfo{year}{2008}),
  \eprint{arXiv:0803.2158 [hep-ex]}.

\end{thebibliography}

\end{document}
%
% ****** End of file apssamp.tex ******